# Affine Transformation Edited and Refined Deep Neural Network for Quantitative Susceptibility Mapping

Zhuang Xiong[1], Yang Gao[1,2], Feng Liu[1], Hongfu Sun[1]

[1]School of Information Technology and Electrical Engineering, University of Queensland, Brisbane, Australia

[2]School of Computer Science and Engineering, Central South University, Changsha, Hunan, China

*Correspondence: Hongfu Sun

**Address**: Room 540, General Purpose South (Building 78), University of Queensland, St Lucia QLD 4072, Australia

**Email**: hongfu.sun@uq.edu.au

**Sponsors:**

HS acknowledges support from the Australian Research Council (DE210101297).

**Abstract**

Deep neural networks have demonstrated great potential in solving dipole inversion for Quantitative Susceptibility Mapping (QSM). However, the performances of most existing deep learning methods drastically degrade with mismatched sequence parameters such as acquisition orientation and spatial resolution. We propose an end-to-end AFfine Transformation Edited and Refined (AFTER) deep neural network for QSM, which is robust against arbitrary acquisition orientation and spatial resolution up to 0.6 mm isotropic at the finest. The AFTER-QSM neural network starts with a forward affine transformation layer, followed by an Unet for dipole inversion, then an inverse affine transformation layer, followed by a Residual Dense Network (RDN) for QSM refinement. Simulation and *in-vivo* experiments demonstrated that the proposed AFTER-QSM network architecture had excellent generalizability. It can successfully reconstruct susceptibility maps from highly oblique and anisotropic scans, leading to the best image quality assessments in simulation tests and suppressed streaking artifacts and noise levels for *in-vivo* experiments compared with other methods. Furthermore, ablation studies showed that the RDN refinement network significantly reduced image blurring and susceptibility underestimation due to affine transformations. In addition, the AFTER-QSM network substantially shortened the reconstruction time from minutes using conventional methods to only a few seconds.

# 1  Introduction

Quantitative Susceptibility Mapping (QSM) is a quantitative contrast mechanism that measures the magnetic susceptibility property of tissue from magnetic field perturbation. Brain QSM has been applied for cerebral vascular injury detection, such as microbleeds and hemorrhage (Chen et al., 2014; Sun et al., 2018), and various neurological disorders related to iron deposition, including Alzheimer's disease (Acosta-Cabronero et al., 2013), Parkinson's disease (Acosta-Cabronero et al., 2016), Huntington's disease (Van Bergen et al., 2016), and multiple sclerosis (Elkady et al., 2018; Elkady et al., 2019). In recent years, deep learning methods have been proposed to solve the ill-posed dipole inversion problem for QSM. For example, QSMnet/QSMnet+ (Jung et al., 2020; Yoon et al., 2018) and DeepQSM (Bollmann et al., 2019) applied the 3-D Unets (Ronneberger et al., 2015) to train the network using *in-vivo* acquired COSMOS (Liu et al., 2009) data and synthetic data, respectively. The xQSM method (Gao et al., 2021) applied octave convolutions to improve the generalization capability of the trained networks. QSMGAN(Chen et al., 2020) showed that the framework of QSMnet could be further enhanced via generative adversarial networks. Apart from dipole inversion, autoQSM (Wei et al., 2019) reconstructed QSM from the unwrapped total field map directly, while iQSM (Gao et al., 2022) computed tissue phase and QSM from the wrapped raw phase images simultaneously. However, these methods have limited generalizability as they require to take local field maps of 1 mm isotropic spatial resolution in pure-axial orientation as their network inputs to achieve optimal performance. Note that it is common to acquire brain QSM scans with a different image resolution than 1 mm isotropic (e.g., $1\times1\times2$ mm$^3$) in an oblique orientation (e.g., AC-PC line). Unrolled deep learning QSM algorithms, such as MoDL-QSM (Feng et al., 2021), LPCNN (Lai et al., 2020), and VaDNI (Polak et al., 2020), have also been proposed to guide the network with the underlying physics-based models. Even though the dipole kernel was incorporated into these unrolled models, performances on testing datasets with dipole kernels substantially different from those in training datasets have not been thoroughly investigated. Besides these supervised learning approaches, FINE (Zhang et al., 2020) edited the network for each testing dataset during inference by tuning the pre-trained model using the model loss. However, this process can be slow, computationally expensive, and limited to low-resolution images due to memory constraints. A more recent method (Oh et al., 2022) applied the adaptive instance normalization technique to account for the image resolution effect, assuming a pure axial acquisition orientation. So far, existing deep learning models for QSM do not generalize robustly to field maps of arbitrary image resolution and acquisition orientation. Because these networks have been optimized to the training datasets of specific acquisition parameters (e.g. 1 mm isotropic resolution in pure axial orientation), the QSM reconstruction results are degraded when acquisition parameters deviate substantially from those of the training datasets. In this work, we propose a new deep network architecture with affine transformation and refinement layers and a comprehensive training dataset with diverse image resolutions and acquisition orientations, to boost the generalizability of the network.

## 2 Methods

### 2.1 Problem Formulation

The relationship between a field-strength normalized local field $\varphi$ and its magnetic susceptibility $\chi$ can be described using the following equation in k-space (Koch et al., 2006):

$$\varphi(\vec{k}) = D(\vec{k}) \cdot \chi(\vec{k}), \tag{1}$$

where $D$ is the unit dipole kernel:

$$D(\vec{k}) = \frac{1}{3} - \frac{(p_x k_x + p_y k_y + p_z k_z)^2}{k_x^2 + k_y^2 + k_z^2}, \tag{2}$$

$[x, y, z]$ are the image-space coordinates; $[k_x, k_y, k_z]$ are the k-space coordinates; $[p_x, p_y, p_z]$ are the vector projections of the subject frame of reference (i.e., field-of-view (FOV)) onto the main magnetic field vector $\vec{B_0}$ (Liu et al., 2009). This vector projection varies with different acquisition orientations.

Given the image voxel size $[v_x, v_y, v_z]$ in millimetre and image matrix size $[M_x, M_y, M_z]$, the k-space coordinates can be calculated as:

$$k_x = \frac{x}{M_x v_x}, \; k_y = \frac{y}{M_y v_y}, \; k_z = \frac{z}{M_z v_z}. \tag{3}$$

When an acquisition is obtained with an isotopic image resolution (e.g., $[v_x, v_y, v_z] = [1,1,1]$ (in mm)) and a pure-axial FOV (i.e., $[p_x, p_y, p_z] = [0, 0, 1]$), dipole kernel in Eq. (2) is simplified to:

$$D(\vec{k}) = \frac{1}{3} - \frac{k_z^2}{k_x^2 + k_y^2 + k_z^2}, \text{ where } k_x = \frac{x}{M_x}, k_y = \frac{y}{M_y}, k_z = \frac{z}{M_z}. \tag{4}$$

Most previous deep learning QSM methods successfully resolved dipole inversion in the specific domain of Eq. (4); however, they failed to generalize to Eq. (2) of arbitrary image resolution and FOV orientation.

## 2.2 Network Design

Here we propose an AFfine Transformation Edited and Refined (AFTER) neural network for QSM dipole inversion. As illustrated in Fig. 1, the AFTER-QSM architecture comprises four steps, of which the first and third steps are two affine transformations, and the second and fourth steps are two deep Convolutional Neural Networks (CNNs):

$$\begin{aligned} \text{step 1:} \quad & \varphi_{\text{lab}} = A \cdot \varphi_{\text{subject}} \\ \text{step 2:} \quad & \chi_{\text{lab}} = \text{Unet}(\varphi_{\text{lab}}) \\ \text{step 3:} \quad & \chi_{\text{subject}} = A^{-1} \cdot \chi_{\text{lab}} \\ \text{step 4:} \quad & \chi_{\text{refined}} = \text{RDN}(\chi_{\text{subject}}) \end{aligned}$$

(5)

The subscripts 'lab' and 'subject' refer to the local field ($\varphi$) or susceptibility ($\chi$) results in their lab frame of reference (i.e., scanner coordinates, $\vec{B_0}$ aligned with z-axis) and subject frame of reference (i.e., prescribed FOV coordinates). $A$ and $A^{-1}$ denote the forward and inverse affine transformations converting between the 'lab' and 'subject' coordinates. Unet (Ronneberger et al., 2015) and RDN (Zhang et al., 2018) ("Residual Dense Network") are two CNN models to be trained jointly. Specifically, in the first step, an input local field map ($\varphi_{\text{subject}}$) of the original resolution in the acquisition subject frame is resampled to an auxiliary local field map ($\varphi_{\text{lab}}$) of isotropic resolution in pure axial orientation (i.e., the z-axis aligns with the scanner $\vec{B_0}$ direction), through an affine transformation $A$. The auxiliary local field map is then fed into a modified 3D Unet, designed to perform dipole inversion for acquisitions of isotropic resolution in pure-axial orientation. The third step reverts the intermediate QSM result ($\chi_{\text{lab}}$) to the original orientation and resolution through the inverse affine transformation $A^{-1}$. Finally, a 3D RDN refinement network is cascaded to remove blurring and susceptibility underestimation induced by interpolation operations in the previous steps.

The 3D forward affine transformation ($A$) is achieved through a 4×4 matrix, combining scaling and rotation:

$$A = \begin{bmatrix} a_{11} & a_{12} & a_{13} & 0 \\ a_{21} & a_{22} & a_{23} & 0 \\ a_{31} & a_{32} & a_{33} & 0 \\ 0 & 0 & 0 & 1 \end{bmatrix}. \tag{6}$$

The 3×3 submatrix on the upper left is the multiplication of the scaling matrix $S$ and the rotation matrix $R$. Here, the rotation matrix is calculated using the quaternion representation, which rotates a unit vector $\vec{l}_2$ (i.e., lab frame coordinate system z-axis) to a given unit vector $\vec{l}_1$ (i.e., subject frame coordinate system z-axis $\vec{B_0}$):

$$R = I + \text{skew}(\vec{l}_1 \times \vec{l}_2) + \frac{\text{skew}(\vec{l}_1 \times \vec{l}_2) \cdot \text{skew}(\vec{l}_1 \times \vec{l}_2)}{1 + \vec{l}_1 \cdot \vec{l}_2}, \tag{7}$$

where $I$ is the identity matrix; *skew* is the operator that generates a skew-symmetric matrix from a given vector $\vec{l} = [i, j, k]$:

$$\text{skew}(\vec{l}) = \begin{bmatrix} 0 & -k & -j \\ -k & 0 & -i \\ -j & i & 0 \end{bmatrix}. \tag{8}$$

Given an arbitrary original acquisition voxel size $[v_x, v_y, v_z]$ (in mm), the scaling matrix $S$ that resamples the native resolution to an isotropic resolution ($v_{\text{iso}}$) is generated as:

$$S = v_{\text{iso}} * \begin{bmatrix} 1/v_x & 0 & 0 \\ 0 & 1/v_y & 0 \\ 0 & 0 & 1/v_z \end{bmatrix}, \tag{9}$$

where $v_{\text{iso}}$ is set to 0.6 mm in this study, which is the finest image resolution of the training dataset. The above 3D forward ($A$) and inverse ($A^{-1}$) affine matrices are used to calculate the image coordinates to be transformed, and the image voxel values are interpolated in the transformed coordinate system using trilinear interpolation.

The tailored Unet, as shown in the middle row of Fig. 1, comprises four encoder and four decoder blocks with channel numbers starting from 16 and doubling after each ConvBlok. The encoder and decoder are connected via concatenations to preserve image structural details and avoid gradients vanishing during backpropagation (He et al., 2016) The modified RDN for susceptibility refinement is detailed in the bottom row of Fig. 1. It consists of four Residual Dense Blocks (RDBs) followed by a global feature fusion (Zhang et al., 2018) to learn global hierarchical features adaptively. Each RDB has a local feature fusion block functioning similarly to the global one. Then, two convolutional layers (kernel sizes: 1×1×1 and 3×3×3) are appended to aggregate all learned feature maps. Convolutional kernels, if not specified, have a size of 3×3×3, with a stride of 1 and a padding of 1. To optimize the networks simultaneously, we formed two loss functions to update the Unet ($loss_1$) and RDN ($loss_2$) models:

$$\begin{aligned} loss_1 &= \left\| \chi_{\text{lab}} - A \cdot \chi_{\text{GT}} \right\|_2 \\ loss_2 &= \left\| \chi_{\text{refined}} - \chi_{\text{GT}} \right\|_2 \end{aligned}, \quad (10)$$

where $\chi_{\text{GT}}$ is the ground truth susceptibility map of native image resolution in its subject coordinate system. $loss_1$ minimizes the difference between predicted QSM and ground truth QSM in the lab frame of reference coordinates with 0.6 mm isotropic resolution. $loss_2$ minimizes the error between the final output and the label QSM in their acquisition coordinates with the original spatial resolution. Adam optimizer was adopted, and the learning rate was empirically set to decay half every 20 epochs from an initial value of $10^{-5}$. This network was trained with a batch size of 2 (due to the GPU memory constraint) for 100 epochs to ensure convergence, using 2 Nvidia Tesla V100 GPUs. The training process took about 4 days.

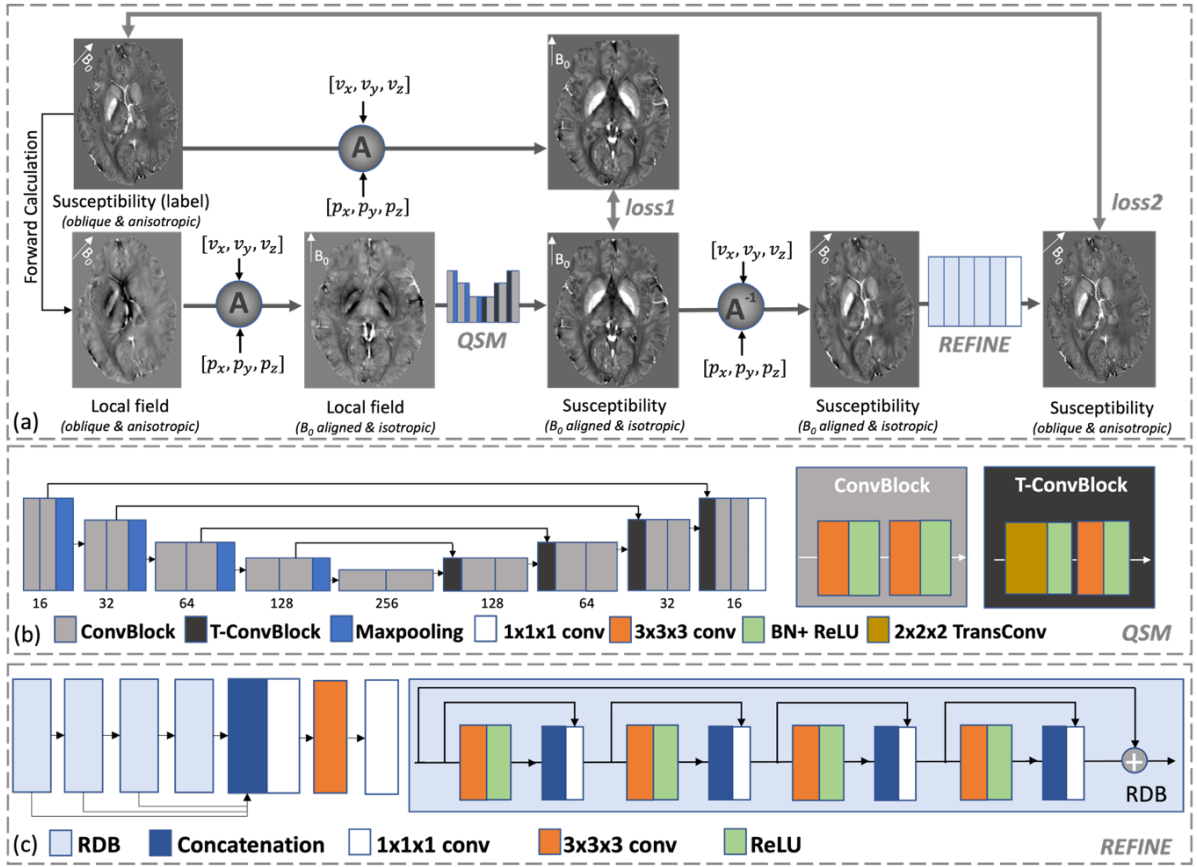

**Figure 1**: (a) Overview of the proposed AFTER-QSM training process. (b) Unet architecture for dipole inversion. Channel numbers are indicated at the bottom of each layer. Convolutions and Transposed

convolutions in ConvBlock and T-ConvBlock have a kernel size of 3x3x3. (c) Residual Dense Network (RDN) architecture for susceptibly refinement. The four Residual Dense Blocks (RDB) outputs are concatenated along the channel dimension.

## 2.3 Training Datasets Preparation

A total of 30 gradient-echo (GRE) scans with full brain coverage acquired at 7 Tesla (matrix size: 334×386×224) of 0.6 mm isotropic resolution from 6 subjects in 5 head orientations were used for generating the training datasets. High-resolution QSM image was reconstructed for each of the 30 GRE scans with a standard pipeline, including background field removal with the RESHARP method (Sun and Wilman, 2014) and dipole inversion using the iLSQR method (Li et al., 2015). First, 3750 QSM patches (matrix size: $96^3$) were cropped from the 30 full-size QSM volumes with a stride of [32, 50, 32] in three dimensions. Second, all patches were zero-padded into a size of $128^3$ to ensure the corners did not overflow the bounds upon rotations. Then, for each QSM patch (i.e., in the scanner coordinates), a random acquisition orientation vector $[p_x, p_y, p_z]$ and a voxel size vector $[v_x, v_y, v_z]$ were assigned to resample the QSM patch in the subject frame coordinates. Finally, local field maps of the patches in the subject coordinates were generated according to Eq. (1). Specifically, each component of the acquisition orientation vector (i.e., $p_x, p_y\ and\ p_z$) was sampled from a standard normal distribution, and then the vector was normalized to a unit vector. The voxel size vector (i.e., $v_x, v_y\ and\ v_z$) was augmented by fixing one random dimension to 0.6 mm, and the other two dimensions were independently sampled from 0.6 mm to 2 mm. We refer to this dataset as the 'mixed' dataset, which was used to train the proposed AFTER-QSM. In addition, to investigate the effect of data augmentation, we generated another dataset called 'pure' dataset from the same 3750 QSM patches but with corresponding local field patches simulated on native 0.6 mm isotropic resolution and pure-axial head orientation (i.e., $[p_x, p_y, p_z] = [0,0,1]$) only.

## 2.4 Testing Datasets Evaluation

We compared the proposed AFTER-QSM neural network with the conventional iLSQR method (Li et al., 2015), three Unet-based deep learning methods, and an unrolling-based deep neural network LPCNN (Lai et al., 2020) on the simulation datasets, and added another single-step conventional method TGV-QSM (Langkammer et al., 2015) to the comparison for *in-vivo* experiments. The three Unet methods are referred to as Pure-Unet, Mixed-Unet, and Affined-

Unet. The Pure-Unet model was trained with the 'pure' dataset (i.e., pure axial and 0.6 mm isotropic resolution). The Mixed-Unet was trained with the 'mixed' dataset (the same as AFTER-QSM). At the same time, the Affined-Unet method applied affine transformations with sinc interpolation (15 voxels width) before and after the pre-trained Pure-Unet model. The intermediate result of AFTER-QSM before the RDN network was also demonstrated in an ablation study to investigate the effect of the susceptibility refinement step.

The brain QSM used in the ablation study (Fig. 2) was obtained from a 1 mm isotropic COSMOS public dataset (Shi et al., 2022). The original COSMOS was downsampled to [1, 1, 2] mm anisotropic resolution and an acquisition orientation vector $\vec{p}$ = [0.71, 0, 0.71] was assigned to simulate a local field map. Another four digital brain phantoms used in the simulation tests, synthesizing different resolution and acquisition angle configurations, were generated from one *in-vivo* high-resolution (i.e., 0.6 mm isotropic) COSMOS map. The simulated QSM image resolutions and FOV orientations are as follows: (#1) pure axial FOV and 0.6 mm isotropic, (#2) titled FOV ($\vec{p}$ = [0.71, 0, 0.71]) and 1 mm isotropic, (#3) pure axial FOV and [0.6, 0.6, 1.6] mm anisotropic, (#4) titled FOV ($\vec{p}$ = [0.71, 0, 0.71]) and [0.6, 0.6, 1.6] mm anisotropic. Peak-Signal-to-Noise Ratio (PSNR) and Structural-Similarity-Index-Measure (SSIM) were computed relative to the ground truth.

In addition, three *in-vivo* acquisitions with different scan parameters (i.e., spatial resolution and FOV orientation) were evaluated to compare different dipole inversion methods. Subject #1 was scanned using a ME-MP2RAGE sequence (Alkemade et al., 2020; Caan et al., 2019; Sun et al., 2020) at a Philips Achieva 7T MRI scanner with a 32-channel head array coil. The inversion times were 670 ms and 3675 ms. A multi-echo readout was added to the second inversion at four echo times (TEs = 3, 11.5, 19, 28.5 ms). Other parameters are: acceleration factor SENSE = 2; FOV = 205×205×164 mm$^3$; voxel size 0.64×0.64×0.7 mm$^3$; sagittal acquisition with a total scan time of 19.53 min. Subject #2 was scanned with a 3T Discovery 750 GE system using a 12-channel head coil. Full brain 3D ME-GRE data were acquired in an axial slab titled 15 degrees to the left shoulder with the following parameters: 8 unipolar readout echoes, 3.4 ms first TE, 3.5 ms echo spacing, 29.8 ms TR, 20-degree flip angle, 1 mm isotropic voxel, 256×256×128 mm$^3$ FOV, ASSET acceleration factor of 2, and a total scan time of 5.9 min. Subject #3 data was acquired at a 7T Siemens whole-body MRI scanner using a multi-echo SWI sequence with a 32-channel head coil. The parameters were: 3D full head coverage using a FOV = 210×210×144 mm$^3$, voxel size = 0.75×0.75×2 mm$^3$, 9 bipolar readouts with first TE = 5.10 ms, echo spacing = 2.04 ms, TR = 24 ms, GRAPPA acceleration factor = 2, first echo flow-compensated, and scan time = 4.71 min.

AFTER-QSM and other dipole inversion methods were compared with the ground truth for simulation studies and with the conventional iLSQR references for *in-vivo* experiments. Correlation and Bland-Altman analysis were performed for the iron-rich deep grey matter regions and a frontal white matter region.

## 3 Results

### 3.1 Ablation Study

Figure 2 shows the AFTRE-QSM reconstruction results of a simulated subject (voxel size: [1, 1, 2] mm, orientation vector: [0.71, 0, 0.71]) in three orthogonal views before and after the RDN refinement network. Note that even though orthogonal views of the brain appear nominal, the head was indeed tilted in the scanner (demonstrated with the cubes in Fig. 2), and FOV was prescribed to align with head orientation. The AFTER-QSM results without the RDN refinement suffered from image blurring and susceptibility underestimation of the vessels. After the RDN refinement, susceptibility maps were noticeably deblurred with structure boundaries sharpened and vessel susceptibilities restored, as evident in the error maps and zoomed-in views. Note that all AFTER-QSM results in the following figures refer to the results after the refinement network.

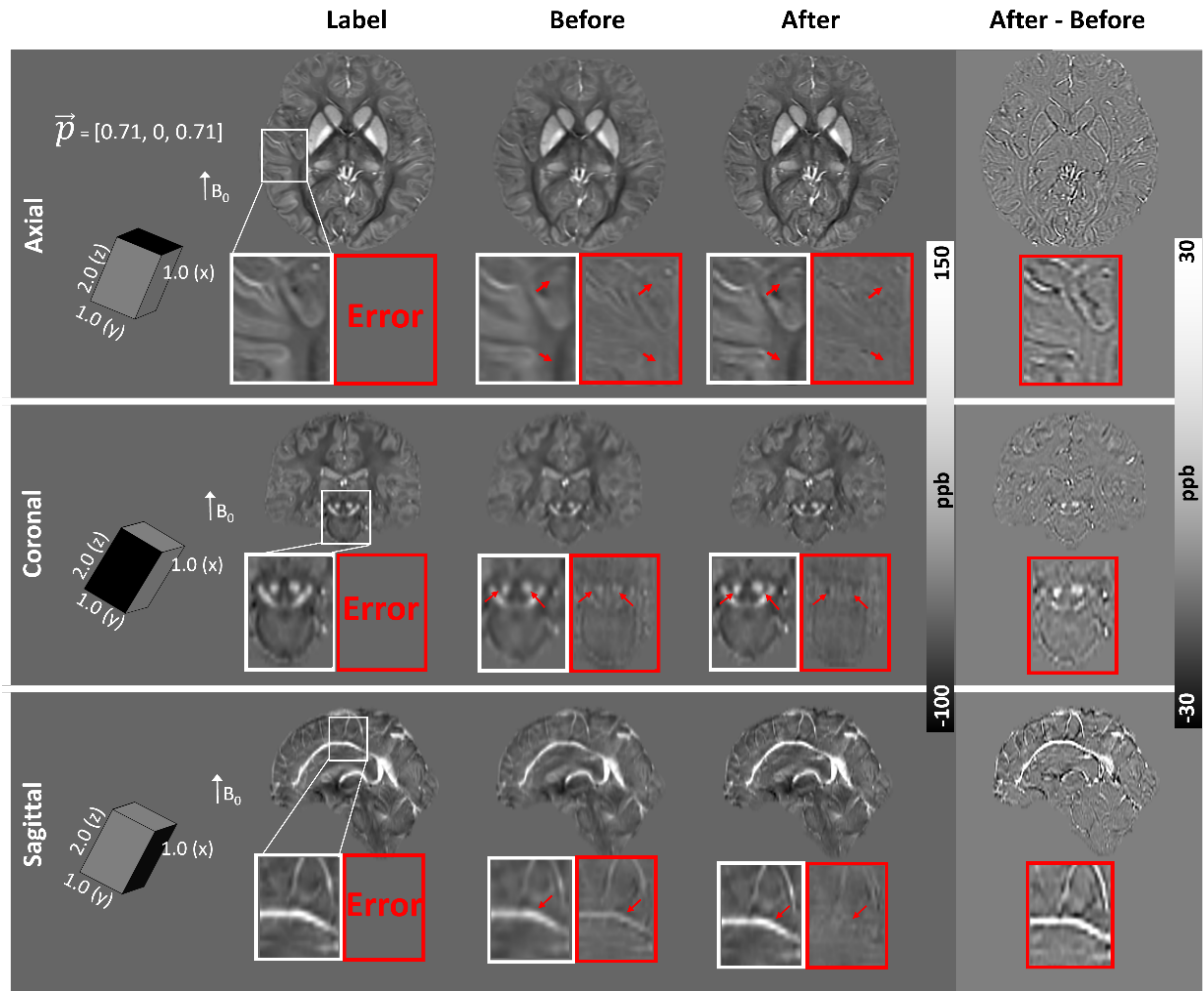

**Figure 2**: Ablation study results demonstrating the effect of the RDN susceptibility refinement network. The AFTER-QSM results before and after the refinement RDN are displayed in the second and third columns, with their differences in the last column. The 3D image volumes are displayed in three orthogonal views. White box regions are zoomed-in to illustrate the deblurring effect after the refinement. The 3D cubes on the left panel describe the acquisition voxel size and FOV angle relative to the main magnetic field direction. The black face of the cubes indicates the direction of the orthogonal views displayed in the same row. Views in coronal and sagittal planes were interpolated and shown in an isotropic resolution for visualization.

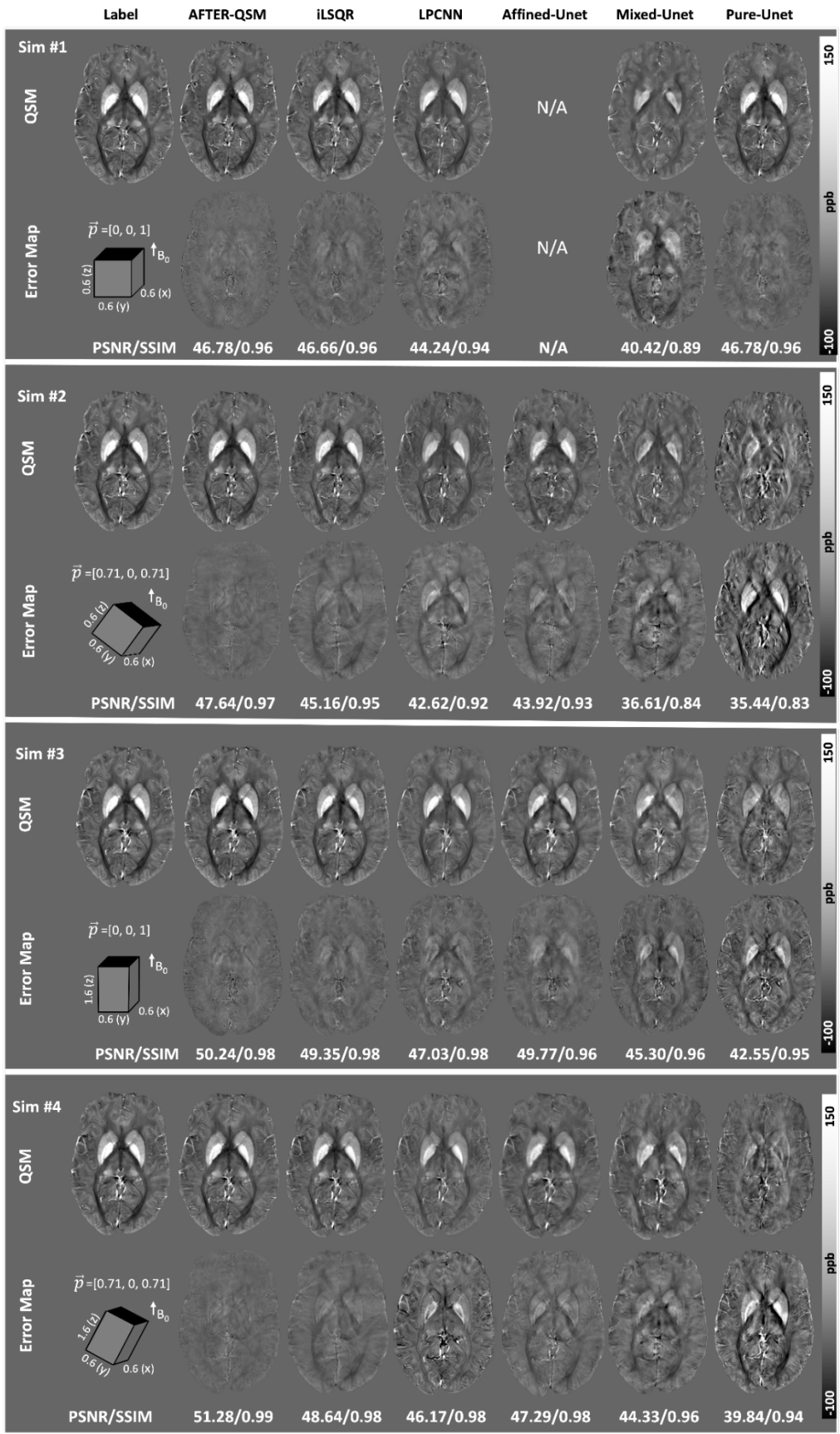

**Figure 3**: Comparison of different QSM methods on four simulated brains of different imaging parameters. (#1) pure-axial FOV, 0.6 mm isotropic, (#2) titled FOV ($\vec{p}$ = [0.71, 0, 0.71]) with 1 mm

isotropic resolution, (#3) pure axial FOV with (0.6, 0.6, 1.6) mm$^3$ anisotropic resolution, (#4) titled FOV ($\vec{p}$ = [0.71, 0, 0.71]) with [0.6, 0.6, 1.6] mm anisotropic resolution. For each simulation test, the top row illustrates the reconstruction results, while the bottom row shows the corresponding error maps. PSNR and SSIM for each image are reported below its error map. The 3D cubes illustrate the acquisition voxel sizes and FOV angles relative to the main magnetic field direction. The black faces of the cubes indicate the directions of the axial views displayed for each simulation case.

### 3.2 Simulation Results

Figure 3 compares different QSM reconstructions for the four simulated digital brain phantoms. For all simulations with different combinations of obliqueness and anisotropy (i.e., #1: isotropic + pure axial, #2: isotropic + oblique, #3: anisotropic + pure axial, and #4: anisotropic + oblique), AFTER-QSM presented the best visually-appealing reconstruction results with the least blurring compared to other methods, which is confirmed by the highest PSNR and SSIM values, followed by the conventional iLSQR method. LPCNN and the two Unet-based approaches showed various degrees of susceptibility underestimation and structural detail loss compared to iLSQR and AFTER-QSM in all cases. Pure-Unet produced the least reliable QSM predictions. Mixed-Unet performed better than Pure-Unet in all cases, except the pure-axial isotropic case (#1). LPCNN performed significantly worse for the oblique angle effect (#2) than the anisotropic resolution effect (#3), and it served the worst when there were both effects (#4). The Affined-Unet was relatively stable in different cases; however, QSM results were significantly blurrier than AFTER-QSM. The experiments were conducted on a desktop computer with a CPU (Intel 12700K), a GPU (RTX 2080Ti of 11GB memory), and a RAM of 64 GB. The reconstruction time for the high-resolution full brain volume (334×386×224) was 37 s for iLSQR, 4.5 s for LPCNN, and 0.7 s for Unet-based methods. For AFTER-QSM, due to the GPU memory constraint, the intermediate result before RDN was cropped along the last dimension (i.e., 224) to 4 overlapping patches (overlapping region size 334×386×16). RDN was performed for each patch sequentially and then resembled. The whole process for AFTER-QSM took 7.4 s.

## 3.3 In-vivo Experiment Results

Consistent with the simulation results, *in-vivo* experiment results in Fig. 4 demonstrate that AFTER-QSM led to comparable results to iLSQR. LPCNN and TGV-QSM output over-smoothed susceptibility maps with reduced susceptibility contrast. Affined-Unet led to blurrier images with suppressed deep grey matter susceptibility contrast than iLSQR and AFTER-QSM. Pure-Unet and Mixed-Unet led to the worst reconstruction results with substantial contrast loss and image artifacts, particularly when the acquisition angles are large (Subjects #1 and #2 in Fig. 4). The AFTER-QSM method achieved consistent results for all three subjects while other deep learning methods' performances were dependent on scan parameters.

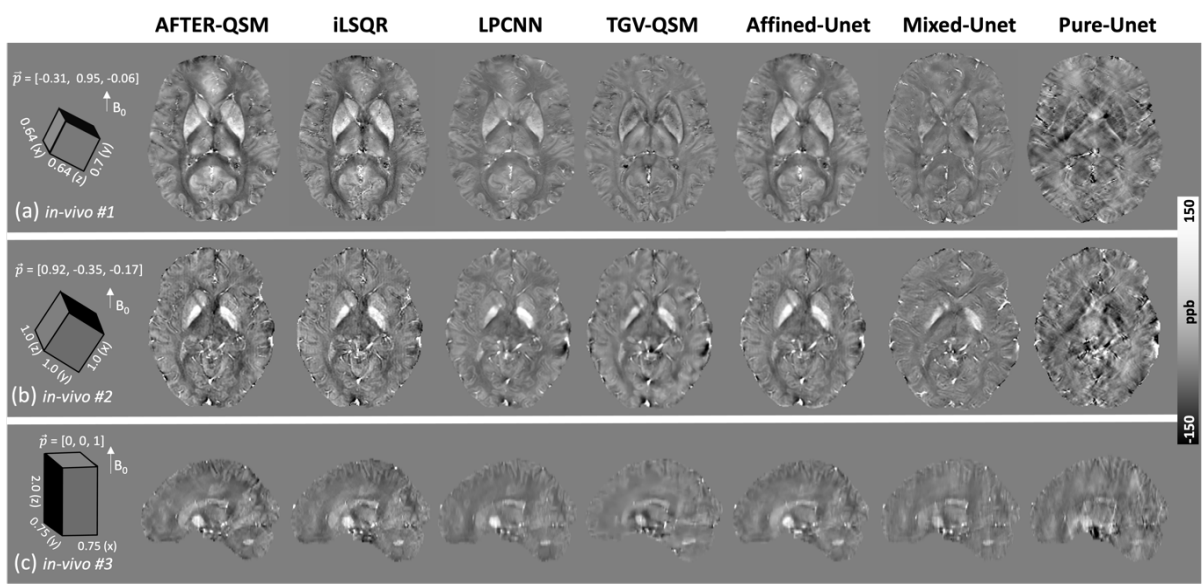

**Figure 4**: Comparison of different QSM methods on three *in-vivo* experiments of different acquisition parameters. FOV orientations and voxel sizes are described using the cubes. (a) A tilted sagittal ($\vec{p}$ = [-0.31, 0.95, -0.06]) acquisition with voxel size = [0.64, 0.64, 0.7] mm. Comparisons are made in the axial view. (b) A titled coronal ($\vec{p}$ = [0.92, -0.17, -0.35]) acquisition with voxel size = [1, 1, 1] mm. Comparisons are made in the axial view. (c) A pure-axial acquisition ($\vec{p}$ = [0, 0, 1]) with voxel size = [0.75, 0.75, 2] mm. Comparisons are made in the sagittal view, which was interpolated and shown in an isotropic resolution for visualization.

## 3.4 Measurements and Correlations

Figure 5 shows the regression and Bland-Altman analysis between the ground truth and reconstructed susceptibility maps in the Caudate Nucleus (CN), Putamen (PU), Globus Pallidus (GP), Thalamus (TH) and Frontal White Matter (FWM) regions (as highlighted in Fig. 5(g)).

For simulation subject #4, the AFTER-QSM method fitted the best to the ground truth with the highest slope and correlation coefficient ($y = 0.92x + 0.01$, $R^2 = 0.98$), followed by Affined-Unet (slope 0.72), Mixed-Unet (slope 0.67) and LPCNN (slope 0.58). The Pure-Unet (slope 0.21) significantly deviated from the ground truth. The reconstruction results of *in-vivo* subject #1 using different methods were compared with the iLSQR reference. AFTER-QSM led to the highest correlation to iLSQR ($y = 1.08x - 0.01$, $R^2 = 0.92$). Mixed-Unet, LPCNN and Affined-Unet showed significant susceptibility underestimation (slopes 0.30, 0.70 and 0.76), while Pure-Unet failed to correlate with iLSQR measurements.

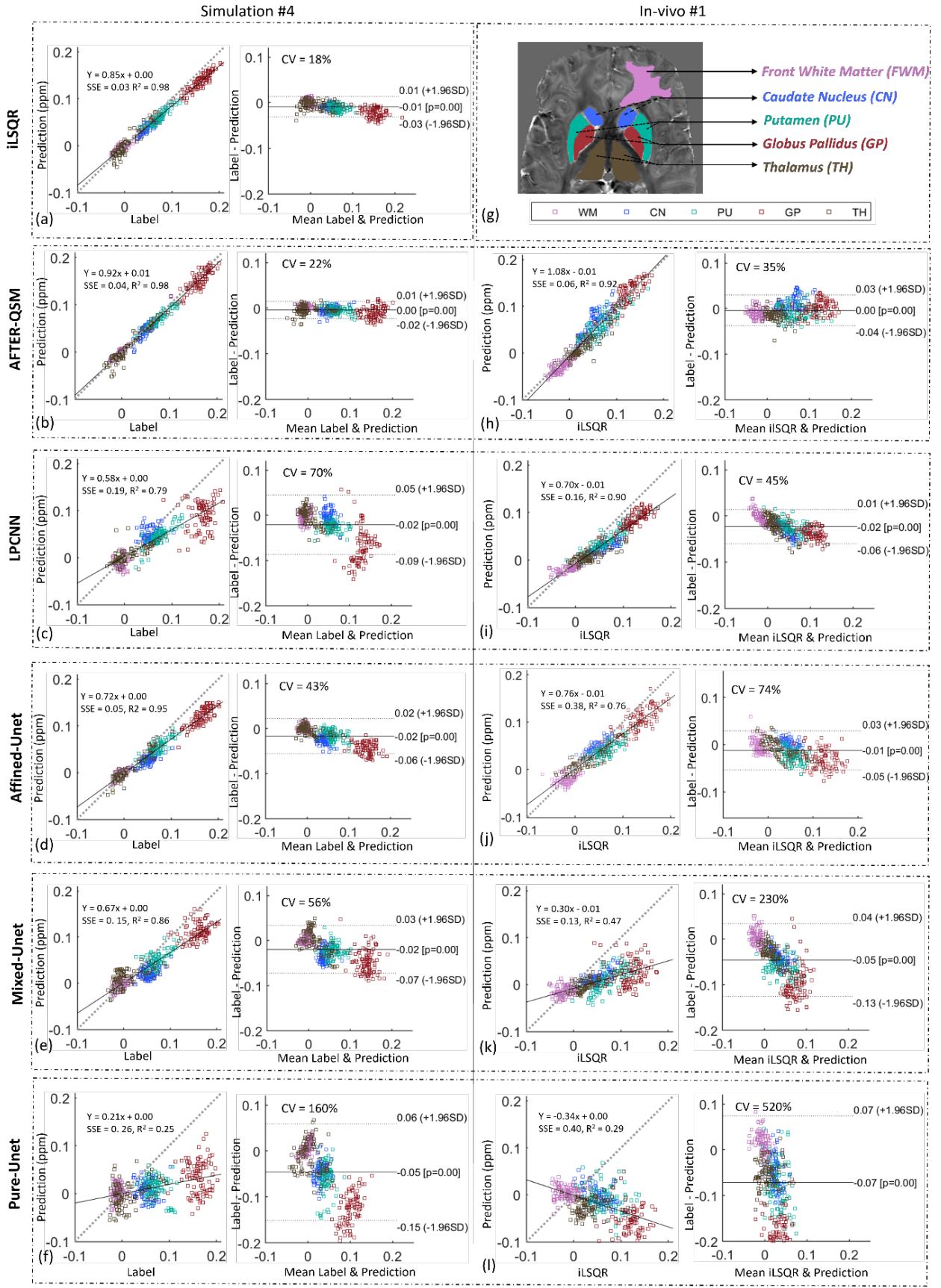

**Figure 5**: Correlation and Bland-Altman plots of deep grey matter and frontal white matter susceptibilities from different methods against the ground truth for simulation subject #4 (a~f) and against the conventional iLSQR method for *in-vivo* subject #1 (h~l). (g) shows the selected deep grey matter and white matter regions for susceptibility measurements. SSE: sum of squared error; $R^2$: coefficient of determination; CV: coefficient of variation; SD: standard deviation; FWM: Frontal White Matter; CN: Caudate Nucleus; PU: Putamen; GP: Globus Pallidus; TH: Thalamus.

## 4 Discussion

In this work, we developed an end-to-end deep learning pipeline for solving ill-posed QSM dipole inversion of arbitrary image resolution (up to 0.6 mm at the finest) and acquisition orientation. The proposed AFTER-QSM method was compared with two traditional methods iLSQR and TGV-QSM, one unrolling-based deep learning method LPCNN, and three Unet-based deep learning methods. The Unet model pre-trained with data of isotropic resolution and pure axial orientation (i.e., Pure-Unet) showed poor generalization on testing datasets of anisotropic and oblique acquisitions, which was slightly improved with Mixed-Unet trained on data with mixed resolutions and orientations. Alternatively, applying affine transformations to the pre-trained Pure-Unet (i.e., Affined-Unet) significantly enhanced the QSM reconstruction, but a considerable degree of blurring and susceptibility underestimation was introduced due to image interpolations from the affine transformations. The Affined-Unet method is similar to the first part of AFTER-QSM. However, the Affined-Unet was trained with pure axial and isotropic resolution, while AFTER-QSM was trained with mixed orientations and resolutions. Furthermore, the AFTER-QSM mitigated the blurring effect and restored the susceptibility contrast with an RDN refinement network at the end of the pipeline. The results showed that AFTER-QSM outperformed other methods in simulation and *in-vivo* experiments.

The RDN refinement network is similar to designs in the image super-resolution tasks (Chen et al., 2018a; Chen et al., 2018b; Zhang et al., 2018). We applied global feature fusion layers (Zhang et al., 2018) to integrate multi-scale features from different levels of RDBs, which is beneficial in solving the problem of gradient vanishing. During network training, the computing graph of the gradients between two deep models was detached. Therefore, Unet and RDN were updated independently with the two loss functions (Eq. (10)). This detachment is more memory efficient and guarantees that Unet is not affected by the subsequent RDN, which accelerates the convergence of training.

Previous works (Feng et al., 2021; Lai et al., 2020) reported that unrolling-based deep learning methods, including LPCNN, were more robust to acquisitions of oblique orientation and anisotropic resolution by incorporating the physics model of the dipole kernel. However, LPCNN was trained using the single-orientation local field map as input and COSMOS from multiple scans as the label, which led to overly smoothing and thus under-estimated results. In addition, the performance of LPCNN was also significantly degraded when the dipole kernel in the testing data was substantially different from those used during network training, as shown in our results.

There are some limitations of this study. First, we chose to upsample all input image resolutions to 0.6 mm isotropic for training the Unet. Even though this improved the network's performance on high-resolution scans, this substantially increased the matrix size of the testing data, resulting in a high memory burden. Second, the refinement network can over-sharpen the images leading to an increased noise level. Future studies will explore the effect of adding different noise levels to the datasets during training. Third, we evaluated the unrolled deep learning method LPCNN using the pre-trained model by the original authors. Re-training LPCNN with the augmented diverse datasets in AFTER-QSM may boost LPCNN's generalization ability. However, it is out of the scope of this technical note.

## 5 Conclusion

We proposed a new AFTER-QSM neural network, which bridges the gap of performance dependency on sequence parameters for exiting deep learning QSM methods. This end-to-end neural network achieved excellent generalizability that can perform dipole inversion of common isotropic or anisotropic resolutions and oblique acquisition orientations in seconds without re-training the model.

## Acknowledgements

HS acknowledges support from the Australian Research Council (DE210101297).